\documentclass[preprint2]{aastex}

\slugcomment{ApJ Letters, 2004, 603, L3-L96}

\shorttitle{Forced oscillations and kHz QPOs}

\shortauthors{Lee, Abramowicz \& Klu\'{z}niak}

\begin{document}


\title{Resonance in Forced Oscillations of an Accretion Disk\\ and
Kilohertz Quasi--Periodic Oscillations}


\author{William H. Lee}
\affil{Instituto de Astronom\'{\i}a, 
Universidad Nacional Aut\'{o}noma de M\'{e}xico, \\ 
Apdo. Postal 70-264, Cd. Universitaria, 
M\'{e}xico D.F. 04510}

\author{Marek A. Abramowicz}
\affil{Astrophysics Department, Chalmers University, S--41296
G\"{o}teborg, Sweden}

\and 

\author{W\l odek Klu\'{z}niak} 
\affil{Institute of Astronomy, Zielona G\'{o}ra University, Lubuska
2, 65--265 Zielona G\'{o}ra, Poland
\\Copernicus Astronomical Center, ul. Bartycka 18, 00-716 Warszawa, Poland}



\begin{abstract}
We have performed numerical simulations of a radially perturbed
``accretion'' torus around a black hole or neutron star and find that
the torus performs radial and vertical motions at the appropriate
epicyclic frequencies. We find clear evidence that vertical motions
are excited in a nonlinear resonance when the applied perturbation is
periodic in time.  The strongest resonant response occurs when the
frequency difference of the two oscillations is equal to one-half the
forcing frequency, precisely as recently observed in the accreting
pulsar, SAX J1808.4-3658, where the observed kHz QPO peak separation
is half the spin frequency of 401 Hz.
\end{abstract}




\section{Introduction}\label{intro}

Millisecond oscillations in the X--ray light curves of systems known
to contain neutron stars or possibly black holes have been observed
for several years with the Rossi X--Ray Timing Explorer (see
\cite{vdk00} for a review). Recently, it has been pointed out that the
centroid frequencies of the corresponding kilohertz quasi periodic
oscillations (kHz QPOs) are in rational ratios of small integers, as
3:2 \citep{ak01,rmmo02,Mc03,abbk03,akklr03}.  This supports
suggestions that a resonance of some kind is responsible for the
observed properties \citep{ka01,ka03,t02}.  In addition, for at
least one system in which coherent pulsations have been detected,
indicating the spin frequency ($\nu_{s}=401$~Hz in SAX J1808.4-3658),
the separation in frequency between the kHz peaks has been recently
reported to be consistent with $\nu_{s}/2$ \citep{wetal03}, implying
that the pulsar is exciting motions in the accretion disk in a
nonlinear fashion \citep{kakls03}.

In this Letter we show that the kHz oscillations detected in SAX
 J1808.4-3658 can be attributed to forcing of epicyclic motions in the
 accretion disk by the 2.5 ms pulsar, which induces resonance
 at selected frequencies. The coupling between the pulsar and the disk
 could be due to the magnetic field, or to some structure on the
 surface of the star.  In other, similar systems, a frequency
 separation equal to the stellar spin frequency is also possible.

\section{Response of a torus to an external perturbation}\label{response}

In order to study the response of an accretion disk to an external
perturbation, we make several simplifying assumptions. The first is to
consider not a full accretion disk but a slender torus in hydrostatic
equilibrium, orbiting a central body of mass $M$. By virtue of the
pressure effects, the rotation curve is not Keplerian, and for all the
cases studied here, we consider constant distributions of specific
angular momentum within the torus, which has low mass and is slender
in the sense that its mass $m \ll M$, and its extension $L \ll R$,
where $R$ is the distance separating it from the mass $M$. We thus
neglect the self--gravity of the torus, which is ellipsoidal in
cross section in a meridional slice. We additionally assume azimuthal
symmetry.

The second assumption is to use a potential for the central mass
$\Phi_{KL}=M[1-\exp(r_{ms}/r)]/r_{ms}$ \citep{kl02}, which reproduces
two features of Einstein's gravity that are relevant in our context:
(1) there is a marginally stable orbit for test masses in circular
orbits at $r_{ms}=6M$, and (2) the vertical and radial epicyclic
frequencies (in the equatorial plane) are given by
$\zeta^{2}(r)=(1/4\pi)^{2}(M/r^{3})\exp(r_{ms}/r)$ and
$\kappa^{2}(r)=(1/4\pi)^{2}(M/r^{3})\exp(r_{ms}/r)(1-r_{ms}/r)$,
respectively, so that their ratio is
$\kappa(r)/\zeta(r)=(1-r_{ms}/r)^{1/2}$, coinciding with the exact
solution in the Schwarzschild metric in general relativity.

In hydrostatic equilibrium, pressure and centrifugal support balance
gravity in the radial direction, and pressure alone balances gravity
vertically.  An ideal equation of state for the fluid closes the
system of equations (we have used an adiabatic index $\gamma=4/3$
throughout). The center of the torus can be conveniently defined as
the locus of maximum density, where the rotation velocity is
Keplerian, and its position is denoted by $r_{0}$. 

Clearly a detailed answer to the general problem being explored here
requires modeling of the full accretion disk, whether it is
geometrically thin or thick. We will explore this in future work. For
the time being, analyzing the behavior of such a slender torus can be
considered analogous to splitting the disk into thin annuli and
investigating the properties of one such ring --- as in analysis of
radiation--induced warping \citep{petterson77,pringle96}. The slender
torus can be thought of as a density enhancement in the accretion
disk, the properties of which might be transmitted and/or amplified
into the overall X--ray light curve observed from the system by an
unspecified mechanism.

\begin{figure}
\plotone{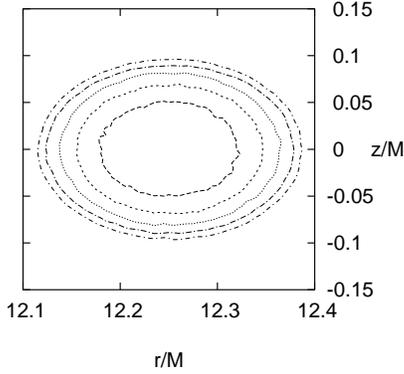}
\caption{Logarithmic density contours in a meridional slice of a
slender torus in hydrostatic equilibrium, spaced every 0.2 dex, with
the lowest one at $\log (\rho/\rho_{\mbox{max}})=-1$. The center of the
torus is at $r_{0}=12.25$~M.
\label{torusIC}}
\end{figure}

The actual tori are constructed from the specified gravitational
potential and equation of state in standard fashion (see
e.g. \citep{ica96,zb86}), and their dynamical behavior (see below) is
followed using a two--dimensional smooth particle hydrodynamics (SPH)
code \citep{monaghan92,lrr02} that uses cylindrical coordinates
$(r,z)$. A cross section (in density) of an equilibrium configuration
thus realized is shown in Fig.~\ref{torusIC}. We now detail the
results obtained from using two different types of perturbations to
such an initial state.

\subsection{Impulsive perturbation}\label{impulsive}

\begin{figure}
\plotone{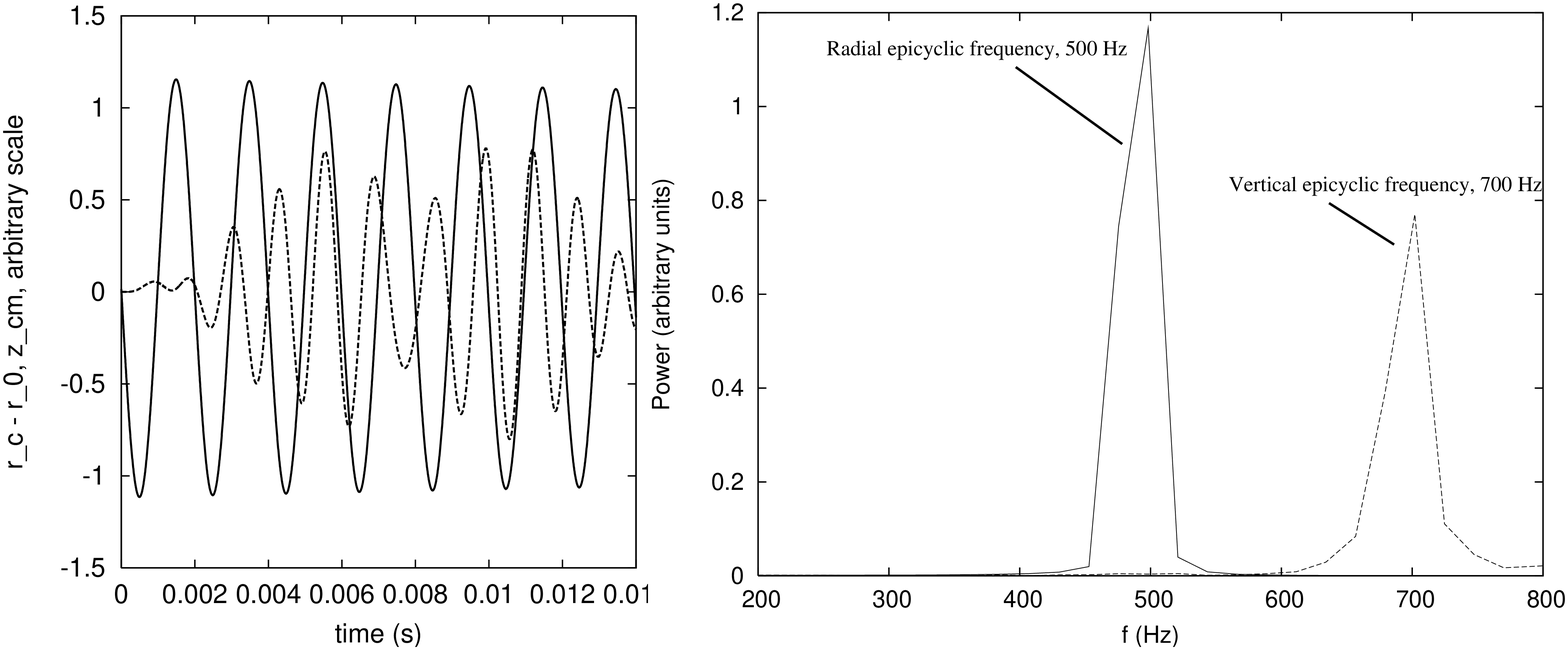}
\caption{Left: Radial (solid line) and vertical (dashed line)
oscillations performed by the center of the torus after an impulsive
radial perturbation at $t=0$. Right: Fourier transforms of the radial
(solid line) and vertical (dashed line) oscillations shown in the left
panel. The two curves have been rescaled so as to fit on the same
graph.
\label{RZimpulsive}}
\end{figure}

As a first example, we consider an impulsive perturbation to the torus,
analogous to that considered recently by \citet{zrf03}. (We note that
these modes might also be relevant for kHz QPOs in black hole systems;
see \cite{rymz03}.)  The equilibrium torus is given a perturbed
velocity field at $t=0$ and allowed to oscillate freely
thereafter. The velocity field of the perturbation is purely radial
(in a cylindrical sense). The kick is small, so that the induced
oscillations in the radial direction have an amplitude that is of the
same order as the extent of the torus itself. The torus performs small
oscillations in the radial direction, primarily at a frequency
consistent with the local epicyclic frequency $\kappa_{0}$, as would
be expected. However, small vertical motions are also induced owing to
pressure coupling between both modes, and these can be clearly seen in
the vertical oscillations of the center of the torus, occurring with
greatest power at a frequency that is consistent with the local
vertical epicyclic frequency, $\zeta_{0}$. Both the radial and
vertical oscillations are shown in Fig.~\ref{RZimpulsive}, along with
their Fourier transforms.

It is thus clear that if radial motions of some kind are present in
the torus, they can induce vertical oscillations, and both of these
occur at the epicylic frequencies for test particles in nearly
circular orbits.

\subsection{Periodic perturbation}\label{periodic}

\begin{figure}
\plotone{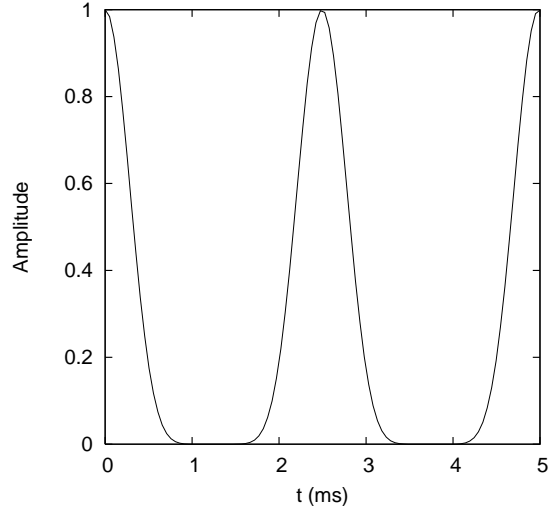}
\caption{Amplitude of the radial perturbation (normalized to the
maximum value) applied to the torus, with a repetition frequency of
$\nu_{s}=401$~Hz.
\label{forcingamp}}
\end{figure}

We now consider a situation in which there is a periodic perturbation
at work, which is external to the disk. The pulsar at the center of
the gravitational potential well is clearly an example of this, and
one would expect it to have an effect on the disk. The magnetic field
of the pulsar, or some deformation on its surface, can perturb the
disk at intervals given by the inverse of the spin period, $\Delta
T=1/\nu_{s}$ (in the case of SAX J1808.4-3658, $\nu_{s}$=401~Hz). We
have considered then a forcing in the radial direction, which
manifests itself through a small, radial acceleration, the magnitude
of which is shown in Fig.~\ref{forcingamp}. It is not purely
sinusoidal but repeats at a fixed interval $\Delta T$, as given
above. The adopted profile of this perturbation deserves some comment:
we have used it to mimic the passage of a brief (compared with the
repetition time), but fairly stronger than average, disturbance in the
accretion disk (e.g., the corresponding polar magnetic field of the
pulsar, sweeping around the disk). Different shapes for this pulse
will be explored elsewhere.

The radial forcing obviously induces a radial oscillation of the
torus. Because of pressure coupling, a vertical oscillation is again
apparent, albeit at a reduced magnitude (compared with the radial
amplitude). The motion of the center of the torus (as defined above)
can be Fourier--analyzed to extract the relevant frequencies. We find
that the radial and vertical motions occur primarily at the local
radial and epicylic frequencies, $\kappa_{0}$ and $\zeta_{0}$
respectively, at $r_{0}$. For a torus orbiting a mass
$M=1.38$M$_{\odot}$, and center at $r_{0}=12.25M=24.8\,$km,
$\kappa_{0}=500$~Hz and $\zeta_{0}=700$~Hz.

The question now is: what effect, if any, does the repetition time
$\Delta T$ of the perturbation have on the power of oscillatory motion
induced in the torus? The radial motion is always driven at a
relatively high amplitude, simply because the perturbation is itself
applied as a radial acceleration. We find, however, that the power of
the vertical motions varies greatly as $\Delta T$ is altered.

\begin{figure}
\plotone{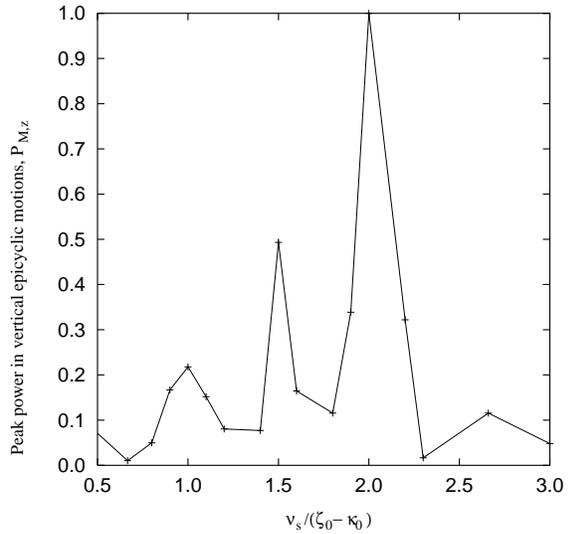}
\caption{Peak power in the vertical epicyclic oscillations induced in
the torus by the periodic perturbation of Fig.~\ref{forcingamp},
as a function of $\nu_{s}/(\zeta_{0}-\kappa_{0})$. Three
peaks at 1.0, 1.5, and 2.0 are clearly visible. For this calculation,
$\nu_{s}$, the pulsar perturbation frequency, varied while
$\zeta_{0}-\kappa_{0}$ was held fixed at 200~Hz.
\label{vertpower}}
\end{figure}

We have kept the initial position of the torus fixed at $r_{0}=12.25M$
and performed over a dozen simulations, differing only in the value of
$\nu_{s}$, and covering a range $100<\nu_{s} \mbox{(Hz)}<600$. For
each one of these, we have computed the peak power $P_{M,z}$ in the
vertical oscillation at $\zeta_{0}$. Since one could equivalently
keep $\nu_{s}$ fixed (as is actually the case in nature) and vary
$r_{0}$ (which would alter the difference $\zeta_{0}-\kappa_{0}$), in
Fig.~\ref{vertpower} we show $P_{M,z}$ as a function of the ratio
$\nu_{s}/(\zeta_{0}-\kappa_{0})$. The response of the torus is clearly
greatest when $\nu_{s}=2 (\zeta_{0}-\kappa_{0})$, as was observed in
the single instance for SAX J1808.4-3658 when two QPO peaks in the kHz
range were seen.

We note that there is also a strong response when $\nu_{s}$ and
$\zeta_{0}-\kappa_{0}$ are in a 1:1 or 3:2 correspondence. The first
of these would allow for the possibility of twin peaks with a
separation of 401~Hz in SAX J1808.4-3658, while the second would imply
a separation of $802/3=267$~Hz. The first option would occur at
$\zeta_{0}$=1054~Hz, $\kappa_{0}$=653~Hz (see also \citet{kakls03}),
and $r_{0}$=9.75M=19.7~km for 1.38M$_{\odot}$. The second would imply
$\zeta_{0}$=832~Hz and $\kappa_{0}$=565~Hz at a radius
$r_{0}$=11.13M=22.5~km for the same mass. Neither has been observed as
yet.

\section{Discussion}\label{disc}

The appearance of twin kHZ QPOs in the millisecond pulsar SAX
J1808.4-3658, with a separation consistent with half the known spin
frequency of the pulsar, strongly indicates that a nonlinear
resonance is at work, coupling the spin to vibrational modes in the
disk.  Using a simple hydrodynamical model, we identify these modes
with the radial and epicyclic oscillations of fluid elements slightly
displaced from exact circular orbits. In this respect, the model is
crucially dependent on the effects of strong gravity, to break the
degeneracy between the orbital and epicyclic frequencies present in
the Newtonian regime.

Under the unique assumption that the pulsar provides a periodic
driving radial force to a slender torus in orbit (which we consider as
a stand--in for a density enhancement in the accretion disk; see
\S~\ref{response}), we show that the response of the torus is greatest
when the spin frequency is twice the difference between the vertical
and radial epicyclic frequencies. We also show that there are other
possibilities for resonant motion, when the above numbers are in a 1:1
or 3:2 correspondence.

For the case of SAX J1808.4-3658, the frequency ratio 700:500=1.4, and
the actual values of the frequencies observed would allow us to fix
the mass of the pulsar at 1.38 solar masses in the Schwarzschild
metric. However, the actual value will be different, as the epicyclic
frequencies for a neutron star rotating at 401 Hz depart from the
Schwarzschild values \citep{kakls03}.

Two further points deserve comment in the context of kHZ QPOs in
systems with a millisecond pulsar whose spin frequency is
known. First, there is an apparent dichotomy in the values of the QPO
peak separation with respect to the spin frequency.  For the ``fast''
rotators, like SAX J1808.4-3658, the separation is half the spin
frequency, $\Delta \nu=\nu_{s}/2$, while for ``slow'' rotators, like XTE
J1807-294 (where $\nu_{s}=190$~Hz), the separation is consistent with
the spin frequency, $\Delta \nu=\nu_{s}$. This could be explained
within the current framework, since for fast rotators, the resonant
point corresponding to $\Delta \nu = \nu_{s}$ is so close to the
neutron star that is it unlikely to be seen (one could say that there
is no room around the pulsar for this mode to occur).  For slow
rotators, both resonances, at $\nu_s$ and $\nu_s/2$, could in principle
be observed, since they would manifest themselves at greater distances
from the neutron star. However, if the modulations in flux generated
in the innermost regions of the accretion disk have a dominant effect
on the overall X--ray light curve, one would then expect to see the
strongest signal at frequency $\nu_s$.

Second, the twin kHz QPOs then need not necessarily occur always at
the same frequencies, as the torus shifts its radial position, but
their separation should nevertheless remain roughly constant (and
equal to $\nu_{s}$ or $\nu_{s}/2$, depending on the system) under this
excitation mechanism, assuming mode locking to occur.  Further timing
observations of millisecond pulsars will no doubt shed light on these
matters.

More generally, our simulation shows directly that a radial perturbation
present in the torus can excite forced oscillation at (other) eigenfrequencies.
This may have implications for mode coupling in black hole accretion disks.

\acknowledgments

Financial support for this work was provided in part by CONACyT
(36632E) and by the Polish Committee for Scientific Research (KBN
grant 2P03D01424). We thank the referee for his comments on the
original version of the manuscript.

\end{document}